\documentstyle[PASJadd,epsbox]{PASJ95}
\newcommand{\vol}{52}
\newcommand{\no}{1}
\newcommand{\lett}{}
\newcommand{\spage}{\bf ??}
\newcommand{\rdate}{1999 September 3}
\newcommand{\adate}{1999 December 22}

\begin{document}

\title{Subaru Observations for the $K$-Band Luminosity Distribution \\
of Galaxies in Clusters near to 3C 324 at $z \sim$ 1.2}

\author{Masaru {\sc Kajisawa}, Toru {\sc Yamada}, Ichi {\sc Tanaka} \\
{\it Astronomical Institute, Tohoku University, Aoba-ku, Sendai 980-8578} \\
{\it E-mail(MK): kajisawa@astr.tohoku.ac.jp } \\
 \\
Toshinori {\sc Maihara}, Fumihide {\sc Iwamuro}, Hiroshi {\sc Terada}, Miwa {\sc Goto},\\
Kentaro {\sc Motohara}, Hirohisa {\sc Tanabe}, Tomoyuki {\sc Taguchi}, Ryuji {\sc Hata} \\
{\it Department of Physics, Faculty of Science, Kyoto University, Sakyo-ku,} \\
{\it Kyoto 606-8502} \\
 \\
Masanori {\sc Iye}, Masatoshi {\sc Imanishi}, Yoshihiro {\sc Chikada} \\
{\it National Astronomical Observatory, 2-21-1, Osawa, Mitaka, Tokyo 181-8588}\\
 \\
Chris {\sc Simpson}, Toshiyuki {\sc Sasaki}, George {\sc Kosugi}, Tomonori {\sc Usuda}, \\
Tomio {\sc Kanzawa} and Tomio {\sc Kurakami} \\
{\it Subaru Telescope, National Astronomical Observatory of Japan,} \\
{\it  650 North Aohoku Place, Hilo, HI 96720, U.S.A. }\\
\vspace{0.5cm}
 }

\abst{
 We investigate the $K$-band luminosity distribution of galaxies
in the region of clusters at $z\sim1.2$ near to the radio galaxy 3C
324. The imaging data were obtained during the commissioning period of
the Subaru telescope. There is a significant excess of the surface
number density of the galaxies with $K =$ 17--20 mag in the region
within $\sim$ 40'' from 3C 324. At this bright end, the measured
luminosity distribution shows a drop, which can be represented by the
exponential cut off of the Schechter-function formula; the best-fitted
value of the characteristic magnitude, $K^{*}$, is $\sim
18.4\pm0.8$. This measurement follows the evolutionary trend of the
$K^*$ of the rich clusters observed at an intermediate redshift, which is
consistent with passive evolution models with a formation redshift $z_f
\gtsim 2$.  At $K \gtsim 20$ mag, however, the excess of the galaxy
surface density in the region of the clusters decreases abruptly,
which may imply that the luminosity function of the cluster galaxies
has a negative slope at the faint end. This may imply strong
luminosity segregation between the inner and outer parts of the clusters,
or some deficit of faint galaxies in the cluster central region of the
cluster. 
}
\kword{galaxies: evolution --- galaxies: formation  --- galaxies: luminosity function, mass function}

\maketitle
\thispagestyle{headings}
%\clearpage
\section{Introduction}

 The luminosity function (LF) is one of the basic probes for studying
galaxy formation. Particularly, the near-infrared $K$-band LF can be
used as a tracer of the galaxy mass distribution since the
near-infrared light of a quiescent galaxy is dominated by low-mass
stars, and can be an approximate measure of the total stellar mass.
Hierarchical structure formation models can be directly tested by
observing the evolution of near-infrared LF over a significant
redshift range (e.g., Kauffmann et al. 1993).  

 Near-infrared observations also have some practical advantages when one
investigates the LF of galaxies at high redshift; it provides a
relatively uniform and unbiased measure of the galaxy luminosity
distribution, since the galaxy luminosity in the $K$-band is much less
sensitive to the on-going star-formation activity, the scale of which
may vary from galaxy to galaxy; also, the $K$-correction factor is
relatively small and nearly independent of the Hubble type, even at
large redshifts.  

 Observing the near-infrared LF in high-redshift-rich clusters may
constrain the history of the evolution of the galaxy mass distribution
in the highest density environment in the Universe. De Propris et
al. (1999) recently investigated the evolution of the $K$-band LF of
galaxies in  rich clusters at $z=$0.1--0.9. The measured LF can be
fitted by the Schechter function, and the behavior of the
characteristic magnitude, $K^*$, along redshift is consistent with those
predicted by the passive-evolution models with a single starburst at
$z=2-3$. Combining these results with the mild evolution of the
mass-to-light ratio of the cluster elliptical galaxies, they concluded
that the assembly of galaxies at the bright end of the LF was largely
completed by $ z\approx 1$.  

  Pushing these studies toward higher redshift constrains the
formation history of galaxies in clusters more strongly. In
a general field environment, Kauffmann and Charlot (1998) claimed a
deficit of red giant elliptical galaxies at $z \gtsim
1$. Franceschini et al. (1998) also argued that the number density of
bright elliptical galaxies significantly decreases above $z \sim
1.3$. Although there is still much debate on these subjects (e.g.,
Totani, Yoshii 1998), it would be interesting to also study the $K$-band LF of
galaxies in rich clusters at $z \gtsim 1$. Observing rich
clusters has an advantage, because the galaxies are at a single distance
and the LF can be evaluated by measuring the galaxy surface-density
excess of the cluster region over the non-cluster region, while it is
still a time-consuming task to spectroscopically measure the distances
to each individual galaxy in the field. The disadvantage is that the
uncertainty of the field correction could seriously affect the
results, especially at the faintest end where the galaxy counts may be
dominated by the foreground/background galaxies. 

 Recently, more than several clusters and cluster candidates at $z
\gtsim 1$ have been discovered (Dickinson 1995; Yamada et al. 1997;
Stanford et al. 1997; Hall, Green 1998; Ben\'\i tez et al. 1999;
Rosati et al. 1999). Most of these objects have been selected or identified
by the surface density excess of the quiescent old galaxy
population. Near-infrared luminosity distributions of the galaxies in
these high-redshift clusters, however, have so far not been studied
intensively. 

 In this paper, we present the $K$-band luminosity distribution of the
galaxies in the clusters at $z \sim 1.2$ near to the radio galaxy 3C 324
using the images obtained with the Subaru telescope. These clusters
are recognized by Kristian et al. (1974) and by Spinrad and Djorgovski
(1984), and firmly identified by Dickinson (1995). The clusters have been
spectroscopically confirmed, and the surface-density excess was
revealed to be due to the superposition of two systems at $z=1.15$
and $z=1.21$ (Dickinson 1997a, b).  The extended X-ray emission, whose
luminosity is comparable to that of the Coma cluster, has been detected
toward the direction of 3C 324 (Dickinson 1997a), which indicates that
at least one of the two systems, probably the $z=1.21$ one in which 3C
324 exists, is a fairly collapsed massive system. Smail and Dickinson
(1995) detected a weak shear pattern in the field that may be produced
by a cluster associated with 3C 324. 
 We describe the observation and the data reduction briefly in section
2. In section 3, we present the obtained $K$-band luminosity function
and the surface number density distribution. Our conclusions and the
discussions are given in the last section. 

\section{Observations and Data Reduction}

 The 3C 324 field was observed at the $K^\prime$ band with the Subaru
telescope equipped with the Cooled Infrared Spectrograph and Camera
for OHS (CISCO, Motohara et al. 1998) on 1999 March 31 and April 1,
during the telescope commissioning period. The detector used was
a 1024 $\times$ 1024 HgCdTe array with a pixel scale of 0.\hspace{-2pt}''116,
which provides a field of view of $\sim$ 2' $\times$ 2'.
 A number of dis-registered images with short exposures
(20 s for each frame) were taken in a circular dither pattern
with a 10'' radius; a series of twelve frames were taken at one
place and the telescope was then moved to the next position. The total
net exposure time was 3000 s. The weather condition was stable during the
observations, and the seeing was $\sim$0.\hspace{-2pt}''8 on March 31 (1600
s) and between 0.\hspace{-2pt}''3 and 0.\hspace{-2pt}''5 on April 1
(1400 s). 

 The data were reduced using the IRAF software package.
There was a variation in the bias level
of CISCO during the observations and the residual pattern is not flat
and makes a 10--20$\%$ discontinuity of the background level in each
frame at the boundary of the quadrants of the array at the central
column of the detector. In the first and the second frames of the
series of the twelve exposures taken at one position, the residual is
much smaller than those in the following other frames. We thus made
the `template' bias
frame, which was to be scaled and subtracted from all frames, by
subtracting the first frame of the seriese averaged over the
observations from the average of the third to twelfth frames. 
 Flat fielding was performed using the frame constructed from our own data
by median stacking of the bias-subtracted images after masking out the
tentatively detected objects. As a check, we compared our
flat-fielding frame with that constructed from many other frames taken
with CISCO by the time of the observations (made by the CISCO team),
and found that their difference was less than 5\%. We indeed performed
the same analysis using this general flat-fielding frame, and found no
systematic difference in the results.  
 After the flat-fielding, by fitting the
background with the 10th-order surface function after masking the
detected objects, we subtracted the sky background and the
remaining slight distortion near to the central column due to the bias
variation and a small pattern of an unfocused image, which was probably
caused by a piece of dust on the camera window, and appeared in the
top-left quadrant of the frame during our observations.
Since the fitted surface was smooth and had a small
gradient over the frame, except for the regions of the dust pattern and
near to the central column, source detection and photometry were little
affected by this procedure. As a further check, we also performed the
median-sky (prepared from the disregistered data frames) subtraction
before removing the spurious patterns by the surface fitting, and found
little difference in the results.  
 The resultant frames were convolved with the Gaussian kernel in order
to  match the FWHM of the stellar images to the worst one (0.\hspace{-2pt}''8).
 They were then co-registered and normalized to be median
stacked. 

 The flux densities of the detected sources were calibrated to those
in the $K$-band by using a star in the list of UKIRT Faint Standards,
FS 27 ($K-K^\prime = -0.01$), observed just after the 3C 324 field at
a similar zenith distance. Since the true-color term of the
telescope and the instrument has not been defined at this stage, and we
do not have infrared colors of the objects, no color correction was
applied. Using the model spectra and the transmission curve of
the CISCO $K^\prime$ filter, we evaluated $K-K^\prime \sim -0.1$ for
an old passively evolving galaxy at $z=1.2$. Many foreground galaxies
may have bluer infrared spectra and the color correction may be
smaller than this. We checked the stability of the results presented
below by artificially shifting the magnitude zero point by 0.1--0.2 mag,
and confirmed that they are little affected by the procedure. 

 We used the SExtractor software (Bertin, Arnouts 1996) to detect
the objects in our image. A detection threshold of $\mu_K =$ 22.4 mag
arcsec$^{-2}$ over 20 connected pixels was used. Photometry was made
with 3'' diameter apertures. We removed the bright objects
with $K < 18$, whose light profile is consistent with the stellar ones from
the final galaxy catalog. We did not make any star/galaxy separation
at the fainter magnitude, but the contribution of the stars was small
at $K > 18$ and at most $\sim 5--10 \%$ (De Propris et al. 1999). 
A total of 146 sources were cataloged. 

 Figure 1 shows the distribution of the detected objects on the
sky. The position angle of the frame is 218$^\circ$. We show the
objects with $K < 20$ mag by the filled circles and those with $K >
20$ by the open ones. In order to evaluate the completeness of the
object detection, we performed a simulation using the IRAF ARTDATA
package. An artificial galaxy with a given apparent magnitude is
generated and added on the observed frame at random coordinates, and
the source-detection procedure with the same threshold was performed
to check whether the artificial object can be detected or not. By
repeating this procedure, we estimated the probability that a galaxy
with a given magnitude is detected.  Each artificial galaxy has a
light profile with parameters randomly selected from the range of
half-light radius between 0.\hspace{-2pt}''1 and 0.\hspace{-2pt}''8
(corresponds to 0.9--6.9
kpc at $z=1.2$ for $H_0=50$ km s$^{-1}$ Mpc$^{-1}$ and $q_0=0.5$) and
that of the axial ratio between  0.3 and 1.0. While the range of the
radius was chosen to represent typical galaxies at $z=1.2$, it covered the
size of the typical field galaxies. Yan et al. (1998) showed that the
size of galaxies with $H=$ 19--23 mag (roughly corresponding to
$K=$ 18--22 mag) is between 0.\hspace{-2pt}''2 and
0.\hspace{-2pt}''6. The seeing effect was
taken into account by convolving the model-galaxy image with the
Gaussian kernel. The result is shown in figure 2. They are
equally-weighted averaged values for the model galaxies with various
sizes and axial ratios.  The detection completeness was $\sim 90$\% at
$K= 21$ mag and still $\sim 70$\% at $K=21.5$ mag; we use the
average value for the disk and de Vaucouleurs profiles in the rest of
this paper. 

  The detection completeness is low for the low surface brightness
galaxies. It may become $\sim 70$\% and $\sim 50$\% at $K=21$ and 21.5
mag, respectively, if we assume an effective radius between
0.\hspace{-2pt}''7 and
0.\hspace{-2pt}''8. This, however, certainly underestimates the true
completeness value, since the faint galaxies have generally smaller
sizes of $\sim$ 0.\hspace{-2pt}''3--0.\hspace{-2pt}''4 at $H=22$ mag
(Yan et al. 1998). 
 
\section{Results}

 Figure 3 shows the differential number counts of galaxies
detected on the entire field of the CISCO $K^\prime$-band image. The
counts were made by 0.5 mag step with one magnitude bin. For
a comparison, those obtained in the general fields taken from various
literatures are also shown. The corrected counts of the frame are
fairly consistent with those of the Hawaii Deep Survey (Cowie et
al. 1994) and in Moustakas et al. (1997) at $K \gtsim 22$ mag, where
the galaxy counts may be dominated by the foreground/background
galaxies.  Bershady et al. (1998) gives systematically higher counts
than others, which may be due to the large-scale structures in the galaxy
distribution and the small area of the observed regions.  

 In figure 4, we show the galaxy surface-density profile in the frame
as a function of the distance from 3C 324 for those objects with $K
=$17--20 mag. The dashed line shows the averaged field surface density
obtained from the literature shown in figure 3. There is a
conspicuous surface-density excess of galaxies within $\sim$ 40''
from 3C 324. The radius corresponds to $\sim 0.35$ Mpc at $z \sim
1.2$. This result is consistent with those presented in figure 3 of
Dickinson (1997b). Dickinson (1997b) also revealed that the galaxies
within 30'' radius from 3C 324 show strong peaks  at $z=1.21$ and
1.15 in the redshift distribution. In the following discussion we do
not distinguish the two systems at $z \sim 1.2$, since the detailed
redshift distribution of the galaxies or the relative population of
them is still unknown. Namely, we discuss the average properties of
the two clusters. Since their redshifts are close, it little affects
the discussion about the absolute luminosity and the color
distributions. In fact, it is a common technique to combine a few
clusters at similar redshifts to reduce the statistical fluctuation of
the galaxy counts.  

 We tentatively divide the observed field into the ``cluster" region,
which is the region within 40'' radius from 3C 324 (shown by the
large circle in Figure 1), and  the adjacent ``outer" region, which is
the remaining region of the frame. The area of the cluster and the
outer regions are 1.433 arcmin$^2$ and 1.981 arcmin$^2$, respectively, and 71
galaxies are in the cluster region. In figure 5, we show the
differential number counts for the two regions separately as well as
the average counts in the literature shown in figure 3, approximated
by the straight line fitted at $K > 17$ mag. In fitting the average
field counts, we put weights by the errors of the values in order to
minimize the effect of the uncertainties in the incompleteness correction
for the data in the literature. 
 Here we plotted the density normalized for the area of the cluster
region so that the readers can see the true numbers of galaxies
detected on our frame in each magnitude bin. As expected from figure
4, the excess of the galaxy surface density of the cluster region is
clearly seen at $K \sim$ 17--20 mag. On the other hand, the counts of
the outer region are similar to the general field counts over the
entire magnitude range, and are likely to be dominated by the
foreground/background field galaxies. 

 At $K \sim$ 20--21 mag, the excess of the surface number density of the
cluster region  decreases very rapidly, and the density even becomes
consistent with those of the field at $K =21$ mag.  The slope of the
counts between $K=19$ mag and 21 mag in the cluster region becomes
fairly flat while the counts of the outer region keeps rising with a
similar slope as in the general fields. 

 The decline of the counts could be due to detection
incompleteness at the crowded region. To check this, we show the
evaluated  detection completeness as a function of the radius from
3C 324 (figure 6). It can be seen that the detection completeness for
the galaxies with $K=$ 20--22 mag only marginally decreases toward 3C
324, except for the innermost bin within 10'' radius where the
effect of the host of 3C 324, which is the brightest cluster galaxy in
the system at $z=1.21$, becomes large and an $\sim 20\%$ decrease of the
completeness is seen.  
 
 Figure 7 shows the obtained luminosity distribution of the galaxies
in the ``cluster" region. For the field correction, the average counts
shown in figure 5 are used. It shows some cut-off at the bright end,
and also drops abruptly toward the fainter magnitude at $K \sim 20$
mag.  We fitted the Schechter function, $\Phi
(L) = \phi^* ( {L \over {L^*}} )^\alpha exp (- {L \over {L^*}} )$,   
 (Schechter 1976), to the bright end of the luminosity distribution
using the data points in the range of 17-20 mag in order to compare
the results with those obtained in a similar manner for the
lower redshift clusters in De Propris et al. (1999) (but there is a
discripancy in that we treat
the differential number counts here, while De Propris et al. (1999)
investigated the cumulative ones). The faint-end slope of the
Schechter function,  
 $\alpha$ = $-0.9$, is assumed as in De Propris et al. (1999). Although
the fitting is far from perfect, the drop toward the brightest end can
be represented by the exponential cut off, as in the Schechter
function. The characteristic magnitude derived in this procedure is
$K^{*} =$ 18.4 $\pm$ 0.8 mag. 

 The apparent rapid drop in the luminosity distribution at $K=$ 20--21
mag seems to be conspicuous. If we assume $\alpha=-0.9$, the expected
number of the {\it cluster} galaxies per magnitude is $\sim$ 9.1 at $K
= 21$ mag, while the observed count between $K=20.5$ and 21.5 mag is
negative after a field correction.  The detection completeness is
still $\sim$ 95--70\% at this depth, and may not be much affected by the
uncertainties in the incompleteness correction. The expected average
number of field galaxies is 22.8 per magnitude at $K = 21$ mag,
while the observed count is 18 and to be 21.6 after the incompleteness
correction. If this behavior is due to a fluctuation of the
foreground/background galaxy counts, there must be a sudden deficit
of about ten galaxies per magnitude just below $\sim 20$ mag and just
inside the 40'' radius from 3C 324.  
 If we assume that there are 9.1 cluster galaxies in the region, the
corresponding number of `observed' field galaxies is 12.5. If the
number density of the field galaxies follows Poisson statistics,
the confidence level of the lower limit (12.5 galaxies) is 99.5\% for
$n=23$. If we assume the presence of 12 and 6 cluster galaxies (9$\pm$3
galaxies), the corresponding number of field galaxies is 9.6 and
15.6, respectively, and the confidence level is 99.98\% and 95.0\%. 
 If we assume that the sum of the cluster and field galaxy counts
approximately follow Poisson statistics, the expected number of
galaxies per magnitude at $K=21$ mag is 31.9 for $\alpha = -0.9$. The
confidence level for the lower limit (21.6) is then 97.7\%. Thus, in fact, the
formal significance of the deficit of cluster galaxies is not very
large. In the real universe, the distribution of galaxies is more
inhomogeneous and the significance may be even lower. However, there
is also no reason that such a relatively rare deficit occurs just in the
`cluster' region where the brighter galaxies are strongly
clustered. We therefore argue that the absence of any number count excess 
at $K = 21$ relative to the expected field galaxy counts is instead an
intrinsic property of the cluster. We can
investigate this further by using the color information for the
galaxies. Indeed, we found four galaxies in the `cluster' region whose
$B-R$ and $R-K$ colors are consistent with those of the galaxies at
$z=1.2$ or higher redshift (Kajisawa et al. 2000),
although we do not see how many of them indeed belong to the
clusters and how many are background galaxies. The true number of
cluster galaxies can only be determined by future complete
spectroscopic surveys or more accurate photometric refshift
measurements. 

 How stable are these results in spite of the non-negligible
uncertainty of the applied field correction ? The galaxy counts of the
``outer" region of the frame may provide another representative field
correction. Although there is a disadvantage that the statistical
uncertainty is large especially at the bright end, due to the small
number of the objects, there is an advantage that the
foreground/background galaxies in the outer region share similar
large-scale structures with those in the adjacent cluster
region. Furthermore, they also share any possible systematic errors in
our data reduction and analysis. At the bright end, $17 < K < 20$, the
counts in the outer region are 10--50 $\%$ smaller than the average
field counts. This does not change the resultant luminosity
distribution of the cluster galaxies very much, since the number of
cluster galaxies is much larger than both the number densities of the
outer region and the average field. On the other hand, at $20 < K <
21.5$, where the deficit of the cluster galaxies is observed, the
number density of the outer region is very similar to that of the
average field (figure 5). Thus, the resultant luminosity distribution
does not change very much at $K \ltsim 21.5$ if we use the counts in
the outer region on the same frame instead of the average field counts
for the field correction. 

 We also examined the surface density profile of faint galaxies
with $K=$ 20--22 mag (figure 8). The raw counts as well as those
corrected using the results shown in figure 6 are plotted. The surface
density within 40'' radius from 3C 324 is even consistent with
those of the field, either the average one or the counts in the
``outer" region in the same frame. We show the expected surface-density
profile within 40'' radius scaled from the excess counts at
$K=$ 17--20 mag assuming the slope of the faint end of LF, ($\alpha$=0,
$-0.9$, $-1.4$), and the characteristic magnitude, $K^* = 18.4$ mag. The
case of $\alpha=-1.4$ cannot be compatible with the observed
data. Even for the case of $\alpha=-0.9$, the observed points are
systematically smaller than the expected counts, although the trend is
somewhat marginal. The observed data seems to be more consistent with the
case of $\alpha=0$, or even that of no surface density excess.  

 It is difficult to constrain the further faint end of the LF below $K
\sim 22$ mag, since the galaxy counts are dominated by the
foreground/background galaxies, even in the ``cluster" region, and the
uncertainty of the incompleteness correction may also greatly affect the
results. We note, however, that the surface density of the
detected objects in the cluster region is systematically higher than
that of the outer region at $K =$ 21.5--23 mag. Some excess of the
cluster galaxies could exist at this magnitude range.

\section{Conclusion and Discussions}

 We presented the luminosity distribution and the surface-density
distribution for the $K$-band selected galaxies in the region of the
clusters at $z=1.15$ and 1.21 near to the radio galaxy 3C 324. While the
bright end of the luminosity distribution can be represented by the
Schechter-function-like exponential cut off with a characteristic
magnitude of $K^* \sim 18.4$ mag, a rapid decrease in the
surface-density excess compared to the average field counts is also seen at
the faint end below $K \sim 20$ mag, $\sim 1.5$ magnitude fainter than
$K^*$.  

  Figure 9 compares the obtained value of $K^{*}$ with those of the
lower-redshift clusters studied by De Propris et al. (1999). Various
lines in the figure show the behavior expected for the no-evolution
and the passive-evolution models with various cosmological parameters
calculated by using GISSEL96 (Bruzual, Charlot 1993). Following De
Propris et al., we used a 0.1 Gyr burst model with a Salpeter IMF and
with solar metallicity. It can be seen that our result follows the
trend of the intermediate-redshift clusters that is consistent with
the passive evolution models with star-formation epoch of $z \gtsim
2$. At $K \ltsim$ 20 mag, the dominant population in the clusters of
3C 324 seems to be old quiescent galaxies which were formed  at
least $\sim 1$ Gyr ago from the observed epoch. 

 The faint-end slope of the optical and near-infrared LF of the nearby
and intermediate-redshift clusters has been studied extensively
(Sandage et al. 1985; Thompson, Gregory 1993; Driver et al. 1994;
Kashikawa et al. 1995; Biviano et al. 1995; Secker, Harris 1996;
Metcalfe et al. 1994; Barger et al. 1996; Smith et al.
1997; Wilson et al. 1997; Driver et al. 1998). Through
near-infrared observations, Barger et al. (1996) give
$\alpha=-1$ for the $K$-band LF of the $z\sim 0.31$ clusters. There
is no such deficit as seen in the 3C 324 clusters at the magnitude
range between $M_K^*$ and $M_K^* + 3$. Wilson et al. (1997) also give
$\alpha=-1$ and $-1.3$ for the $I$-band LF of the two clusters at
$z\sim 0.2$, respectively. Although there are some unevenness within a
factor of two, no rapid drop is seen between $m_I^*$ and $m_I^*+4$.  

 There is evidence that the optical LF of the clusters may be bimodal
and better fitted by the combination of a Gaussian distribution for
the bright (giant) galaxies and the Schechter function for the faint
(dwarf) galaxies rather than by the single Schechter function (Sandage
et al. 1985; Thompson, Gregory 1993; Kashikawa et al. 1995; Biviano
et al. 1995; Secker, Harris 1996; Metcalfe et al. 1994). Indeed,
both the $B$-band and $R$-band LFs of the Coma cluster show some
`gap' at $\sim 1$ mag fainter than the peak magnitude of the bright
population (Biviano et al. 1995; Secker, Harris 1996). In the $R$
band, the characteristic magnitude of the `faint' population is $\sim
1$--$1.5$ mag fainter than the peak magnitude of the `bright' one
(Secker, Harris 1996). The shape of the $H$-band luminosity function of
the Coma cluster (De Propris et al. 1998) is also consistent with
these results.   

 On the other hand, a deficit of the galaxies is seen at $\sim$
1.5--2.5 mag fainter than the $K^*$ in the 3C 324 clusters. Although
there is still an uncertainty of $\sim 1$ mag in the value of the $K^*$,
the deficit of the galaxy seems to occur at the magnitude range where
the dwarf population already begins to dominate in the Coma cluster.
How can we interpret this deficit of the faint galaxies at $K \sim 21$
in the 3C 324 clusters if it is not just a statistical fluctuation of
the foreground/background galaxies and an intrinsic property of the
cluster?  

  It may be due to luminosity segregation between the inner and outer
radius from 3C 324.  The physical diameter of the ``cluster region" (R
$<$ 40'') studied in this paper is $\sim 0.7$ Mpc, and thus should be
considered to be a central part of the cluster. Driver et
al. (1998) have shown that the dwarf galaxies ($M_R > M_R^*+3$) are less
concentrated than the luminous galaxies in the clusters with
Bautz--Morgan type III, namely, irregular less-concentrated
systems. The rich clusters at high redshift may still be dynamically
young and share the properties of irregular clusters seen in the local
universe. If the faint galaxies in the 3C 324 clusters are more
populous in the outskirt region, or distributed rather flat over the
scale of a few Mpc, the excess of the surface density can be easily hidden
by the numerous foreground/background galaxies. 

 Another possibility is the intrinsic deficiency of the faint galaxy
population in the cluster(s). There is a model involving a late formation of
dwarf galaxies.  Kepner et al. (1997) have shown that the UV background
radiation of $J_\nu$ = 10$^{-21}$ erg s$^{-1}$ cm$^{-2}$ Hz$^{-1}$ at
the Lyman limit at $z=3$ and evolves as $(1+z)^4$ can prevent a
baryonic collapse in the dark-matter halo with a circular velocity of
$\ltsim 30$ km s$^{-1}$ by $z\sim1.5$, although the Hubble Deep
Field observation did not prove the presence of many `bursting dwarf'
populations (Ferguson, Babul 1998). The circular velocity, $\sim 30$
km s$^{-1}$, corresponds to the dynamical mass of $\sim 10^9 M_\odot$,
or may be $L \sim 0.01$ $L^*$ and much less luminous than the
$K \sim 21$ mag object ($\sim 0.1$ $L^*$). In a rich cluster environment
at a high redshift, however, the limiting mass could be larger, since
there may be large extra contribution of UV radiation by the massive
stars rapidly formed in the primordial giant elliptical galaxies,
which constitute the bright end of the LF at the observed epoch, in
addition to the general background field, which may be due to quasars.

 The results presented in the paper are just for one region and the
superposition of the two distinct systems could make the situation
more complex. Clearly, it is important to extend the study of the LF
to other high-redshift clusters as well as to constrain more firmly
the faint-end of the LF in the intermediate-redshift clusters. \\
\vspace{0.5cm}

%*********** ACKNOWLEDGES
We would like to thank the referee, Mark Dickinson for his invaluable
comments. The present result is indebted to all the members of the Subaru
Observatory, NAOJ, Japan. This research was supported by grants-in-aid
for scientific research of the Japanese Ministry of Education,
Science, Sports and Culture (08740181, 09740168). This work was also
supported by the Foundation for the Promotion of Astronomy of Japan. 
The Image Reduction and Analysis Facility (IRAF) used in this paper is
distributed by National Optical Astronomy Observatories, operated by
the Association of Universities for Research in Astronomy, Inc., under
contact to the National Science Foundation.

%\clearpage

\section*{References}
\small

\re
Barger A. J., 
Arag\'on-Salamanca A., Ellis R. S., Couch W. J., Smail I., Sharples R. 
M. 1996, MNRAS 279, 1 

\re
Ben\'\i tez N., Broadhurst T., Rosati P., Courbin F., Squires G.,
Lidman C., Magain P. 1999, ApJ 527, 31

\re
Bershady M. 
A., Lowenthal J. D., Koo D. C. 1998, ApJ 505 50 

\re
Bertin E., Arnouts S. 1996, A\&AS 117, 393 

\re
Biviano A., Durret F., 
Gerbal D., Le Fevre O., Lobo C., Mazure A., Slezak E. 1995, A\&AS 
297, 610 

\re
Bruzual A. G., Charlot S.  1993, ApJ 405, 538 

\re
Cowie L. L., Gardner J. 
P., Hu E. M., Songaila A., Hodapp K. W., Wainscoat R. J. 1994, ApJ 
434, 114 

\re
De Propris R., Eisenhardt P. R., Stanford S. 
A., Dickinson M.  1998, ApJ 503, L45 

\re
De Propris R., Stanford 
S. A., Eisenhardt P. R., Dickinson M., Elston R. 1999, AJ
submitted (astro-ph/9905137)

\re
Dickinson M. 1995, ASP Conf. 
Ser. 86, 283 

\re
Dickinson M. 1997a, in HST and the High Redshift Universe, ed
N. R. Tanvir, A. Arag\'on-Salamanca, J.V. Wall (World
Scientific, Singapore) p207

\re
Dickinson M. 1997b, in The Early 
Universe with the VLT, ed J. Bergeron (Springer, Berlin)
p274

\re
Driver S. P., Phillipps 
S., Davies J. I., Morgan I., Disney M. J. 1994, MNRAS 268, 393 

\re
Driver S. 
P., Couch W. J., Phillipps S.  1998, MNRAS 301, 369 

\re
Ferguson H. C., Babul A. 1998, MNRAS 296, 585 

\re
Franceschini A., 
Silva L., Fasano G., Granato L., Bressan A., Arnouts S., Danese L.
1998, ApJ 506, 600 

\re
Hall P. B., Green R. 
F. 1998, ApJ 507, 558 

\re
Kashikawa N., 
Shimasaku K., Yagi M., Yasuda N., Doi M., Okamura S., 
Sekiguchi M.  1995, ApJ 452, L99 

\re
Kauffmann G., 
Charlot S.  1998, MNRAS 297, L23 

\re
Kauffmann G., White S. D. M., Guiderdoni B. 1993, MNRAS 264, 201 

\re
Kepner J. V., Babul A., Spergel D. N. 1997, ApJ 487, 61

\re
Kristian J., 
Sandage A., Katem B. 1974, ApJ 191, 43 

\re
McLeod B. A., Bernstein 
G. M., Rieke M. J., Tollestrup E. V., Fazio G. G. 1995, ApJS 96, 117 

\re
Metcalfe N., 
Godwin J. G., Peach J. V. 1994, MNRAS 267, 431 

\re
Minezaki T., Kobayashi Y., Yoshii Y., Peterson B. A. 1998, ApJ 
494, 111 

\re
Motohara K., Maihara T., Iwamuro F., Oya S., Imanishi M., Terada H., 
Goto M., Iwai J., et al. 1998,  Proc. SPIE 3354, 659 

\re
Moustakas L. A., 
Davis M., Graham J. R., Silk J., Peterson B. A., Yoshii Y. 1997, 
ApJ 475, 445 

\re
Rosati P., Stanford S. A., 
Eisenhardt P. R., Elston R., Spinrad H., Stern D., Dey A. 1999, 
AJ 118, 76

\re
Sandage 
A., Binggeli B., Tammann G. A. 1985, AJ 90, 1759 

\re
Secker J., Harris 
W. E. 1996, ApJ 469, 623 

\re
Schechter P. 1976, ApJ 
203, 297 

\re
Smail I., 
Dickinson M.  1995, ApJ 455, L99 

\re
Smith R. M., 
Driver S. P., Phillipps S.  1997, MNRAS 287, 415 

\re
Spinrad H., 
Djorgovski S. 1984, ApJ 280, L9 

\re
Stanford S. A., Elston R., Eisenhardt P. R., Spinrad H., Stern D., Dey A. 1997, AJ 114, 2232

\re
Thompson L. A., 
Gregory S. A. 1993, AJ 106, 2197 

\re
Totani T., Yoshii 
Y.  1998, ApJ 501, L177 

%Whitmore, B. 
%C., Gilmore, D. M. \& Jones, C.  1993, \apj, 407, 489 

\re
Wilson G. 
, Smail I., Ellis R. S., Couch W. J. 1997, MNRAS 284, 915 

\re
Yamada T., Tanaka I., 
Arag\'on-Salamanca A., Kodama T., Ohta K., Arimoto N.  1997, ApJ 
487, L125 

\re
Yan L., McCarthy P. J., Storrie-Lombardi L. J., Weymann R. J. 1998,
ApJ 503, L19

\clearpage

% Fig.1
\begin{figure*}[p]
\begin{center}
   \epsfile{file=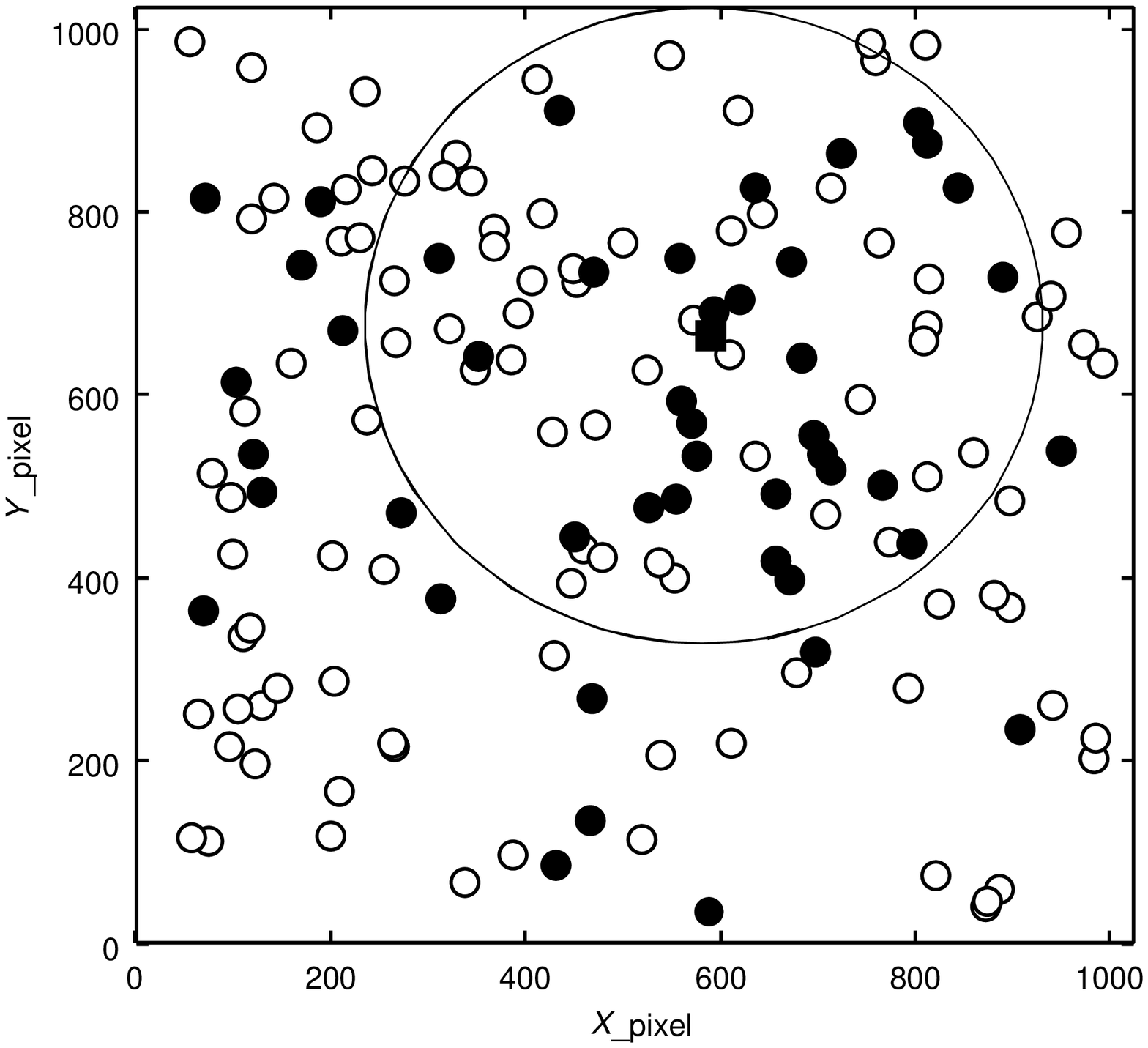,scale=0.65}
 \end{center}
\footnotesize Fig.\ 1.\ Distribution of the 146 detected objects on
the sky. The position
angle of the figure is 218$^\circ$. 3C 324 is shown by the filled
square. The objects with $K < 20$ are shown by the filled circles and
those with $K > 20$ by the open circles. The large circle indicates
the ``cluster" region within 40'' from 3C 324. 
\end{figure*}

\begin{figure*}[p]
\begin{center}
   \epsfile{file=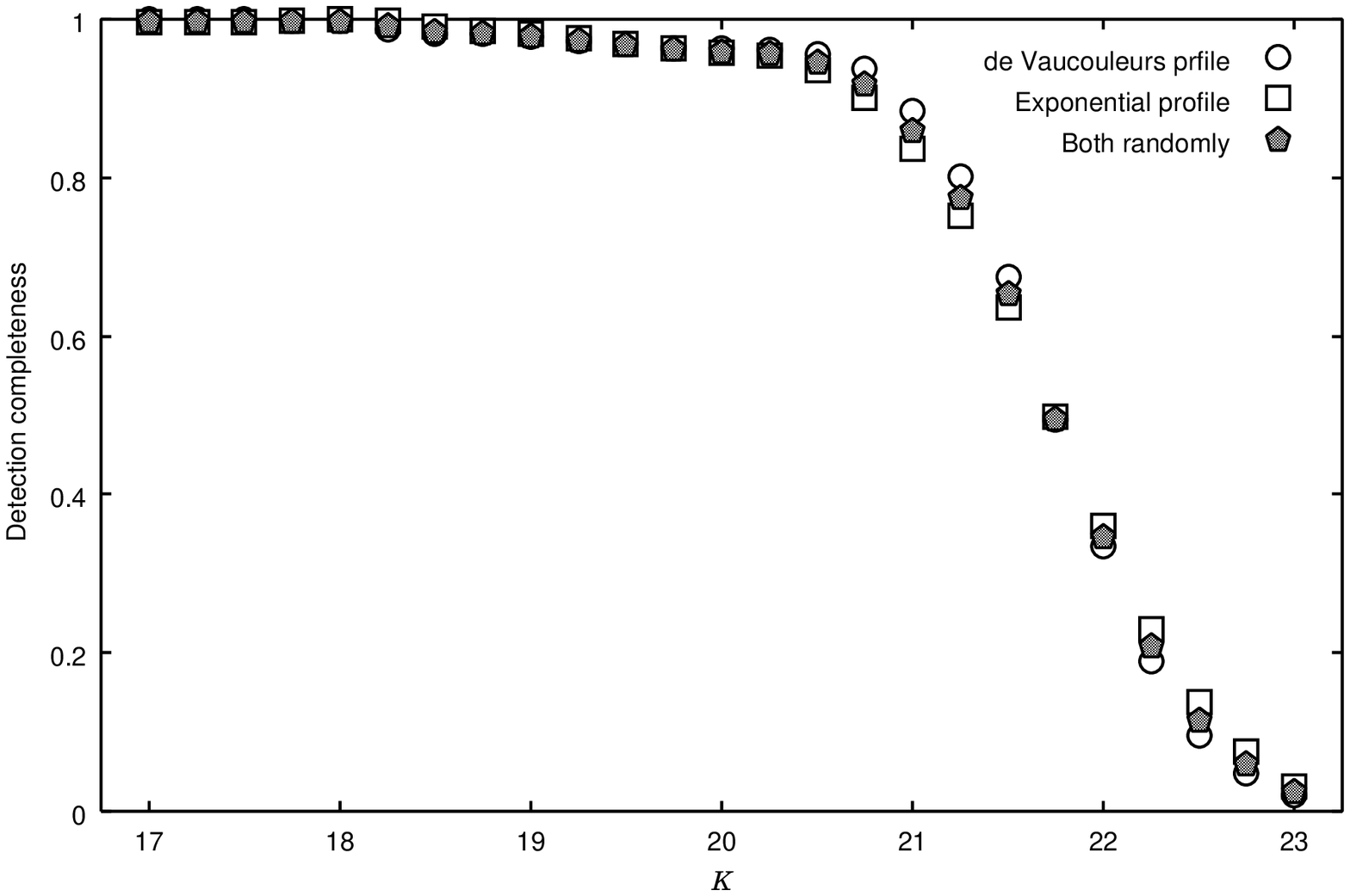,scale=0.65}
 \end{center}
\footnotesize Fig.\ 2.\ Detection completeness as a function of the
apparent magnitude derived from the simulation (see text). Each
symbol represents the adopted profiles of the artificial galaxies; de
Vaucouleurs profile (open circle), exponential profile (open square),
average of both profiles (shaded pentagon), respectively.
\end{figure*}

\begin{figure*}[p]
\begin{center}
   \epsfile{file=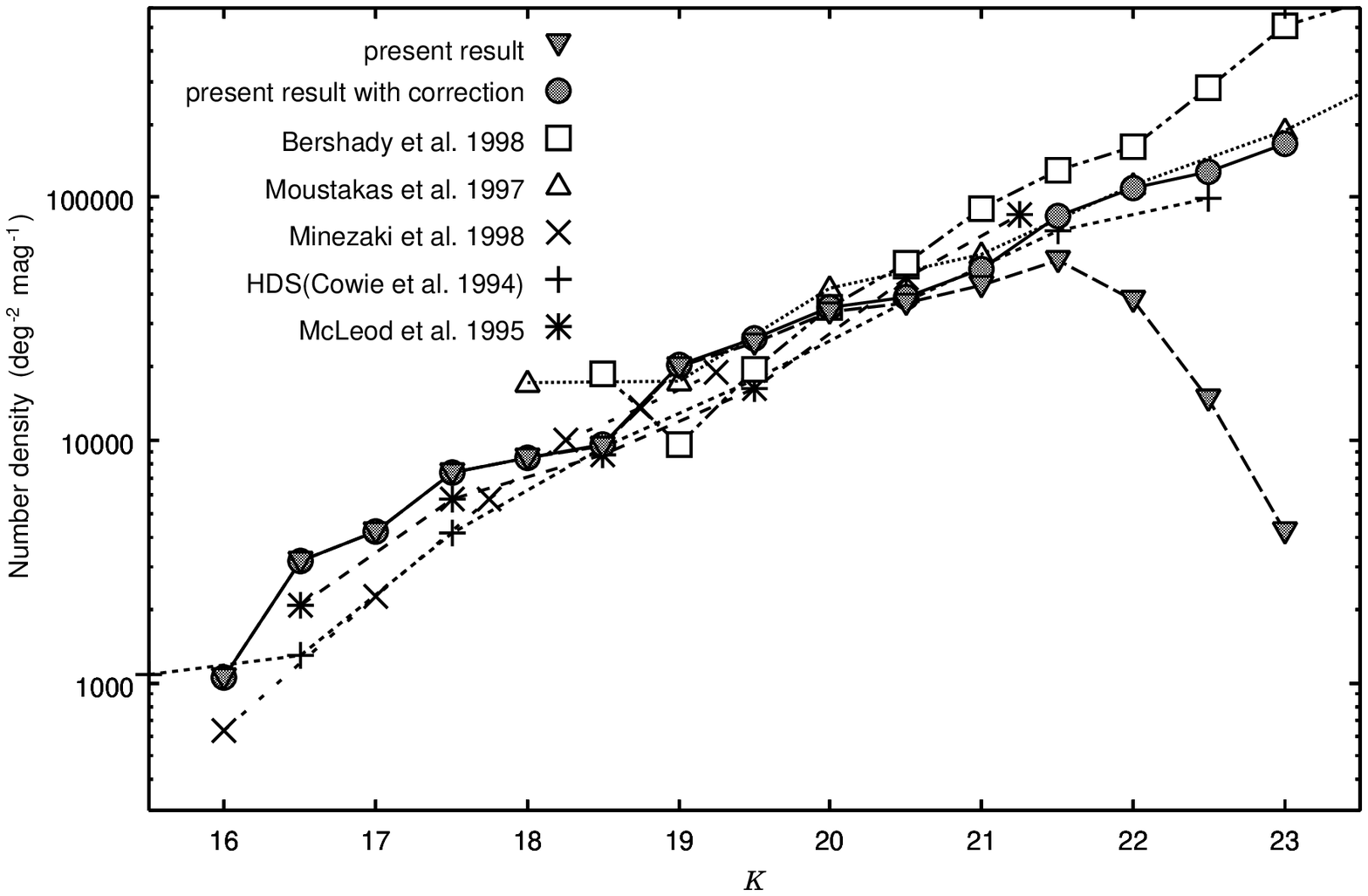,scale=0.65}
 \end{center}
\footnotesize Fig.\ 3.\ $K$-band number counts of the galaxies in the 3C 324 region.
 The observed counts (shaded triangles) as well as the
incompleteness-corrected counts (shaded circles) are shown. The galaxy
counts in the general field taken from literatures are also plotted. 
\end{figure*}

\begin{figure*}[p]
\begin{center}
   \epsfile{file=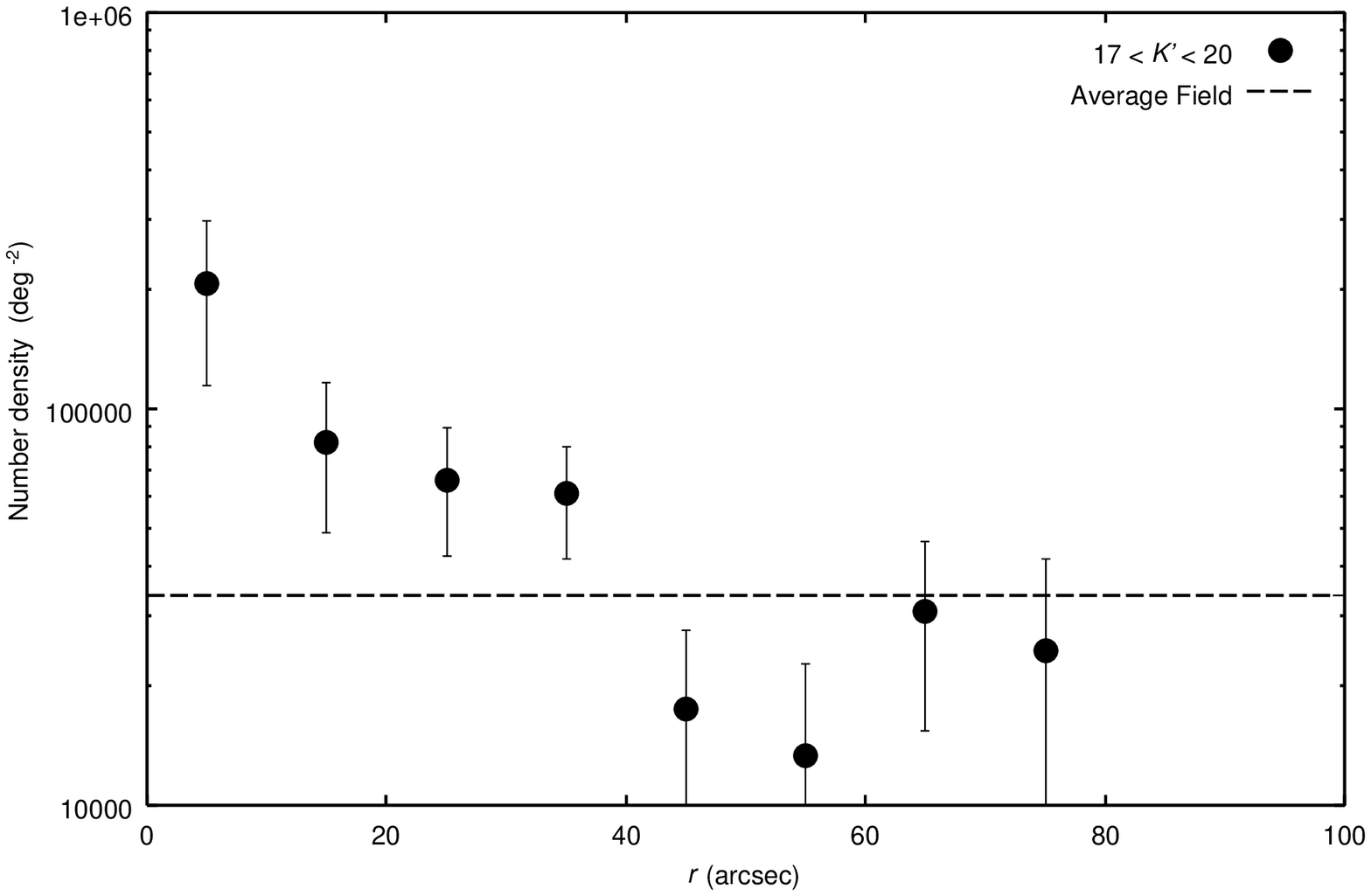,scale=0.65}
 \end{center}
\footnotesize Fig.\ 4.\ Surface number density profile of the galaxies with 17 $< K <$ 20
as a function of the radius from 3C 324.  The dashed line represents the
averaged surface number densities in the field shown in figure 2. The
error-bars represent the square root of the number of detected
galaxies.  
\end{figure*}

\begin{figure*}[p]
\begin{center}
   \epsfile{file=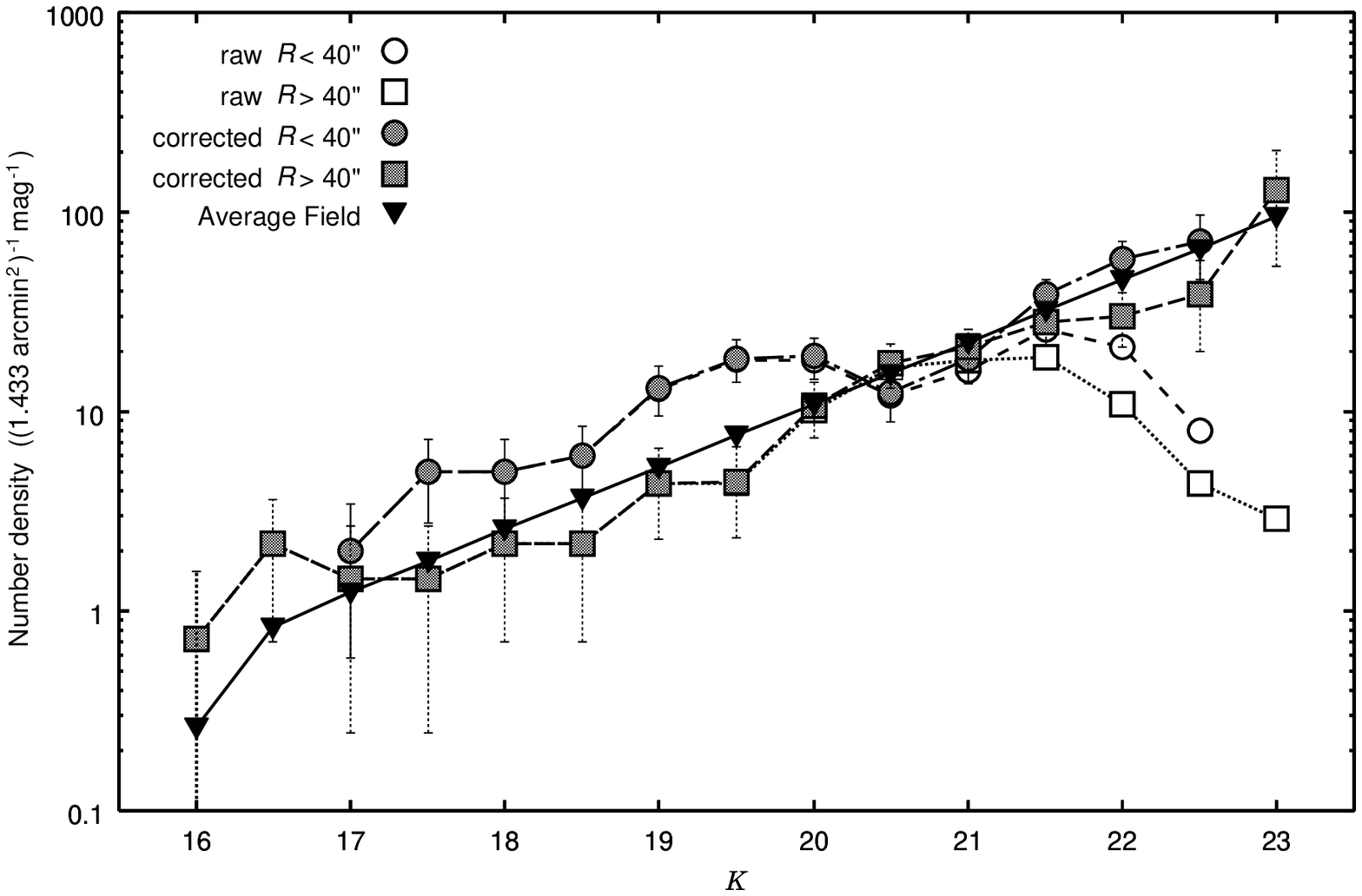,scale=0.65}
 \end{center}
\footnotesize Fig.\ 5.\ Corrected (shaded symbol) and raw (open symbol) $K$-band number
counts of the ``cluster" and the ``outer" regions. The filled
triangles show the approximate average field counts.  The number
density is normalized by the area of the ``cluster'' region. 
\end{figure*}

\begin{figure*}[p]
\begin{center}
   \epsfile{file=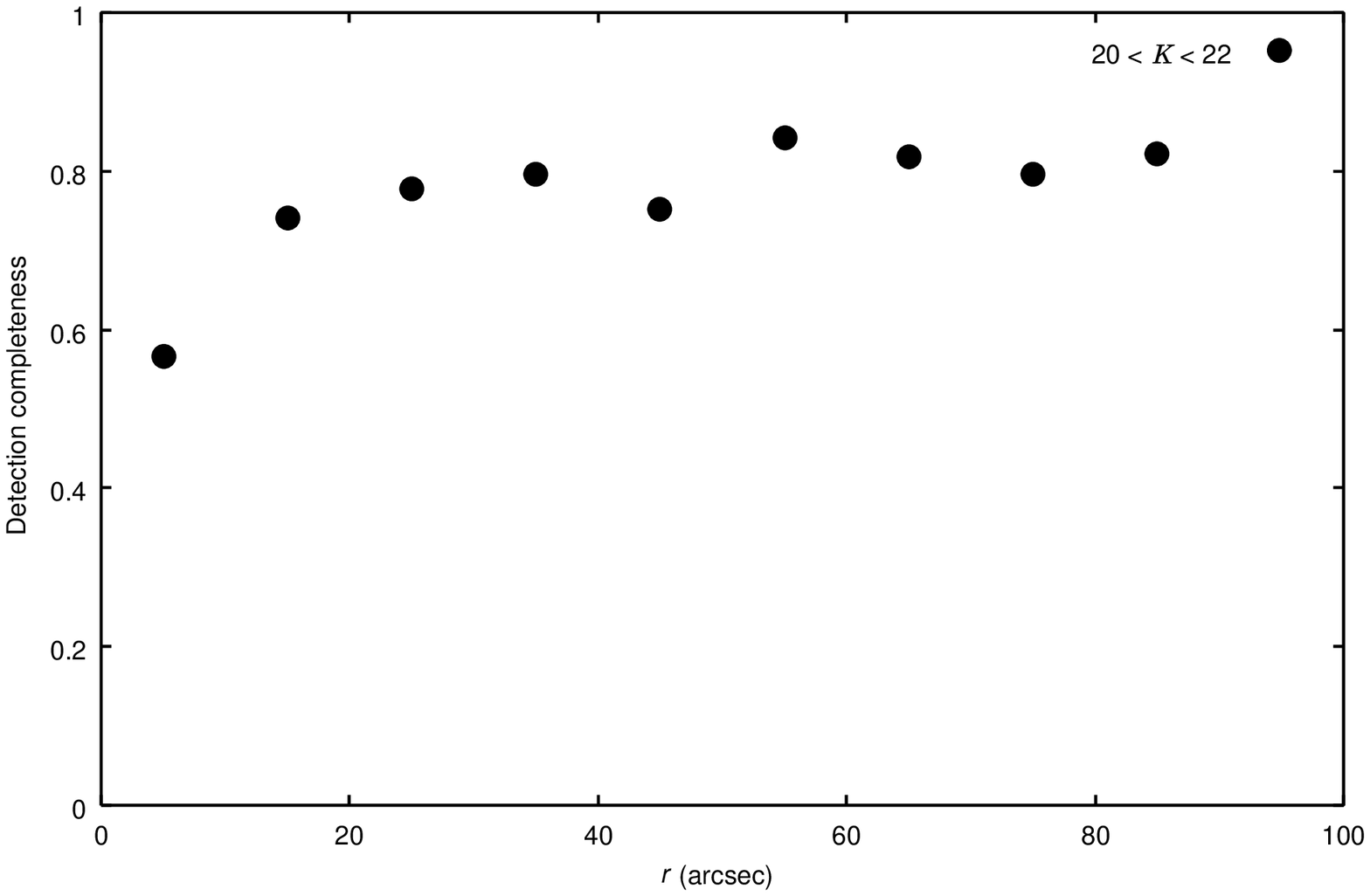,scale=0.65}
 \end{center}
\footnotesize Fig.\ 6.\ Detection completeness as a function of the distance from the 3C 324 for the galaxies with $20 < K < 22$. 
%The error-bars represents the square root of the number of the artificial galaxies detected in each region.
\end{figure*}

\begin{figure*}[p]
\begin{center}
   \epsfile{file=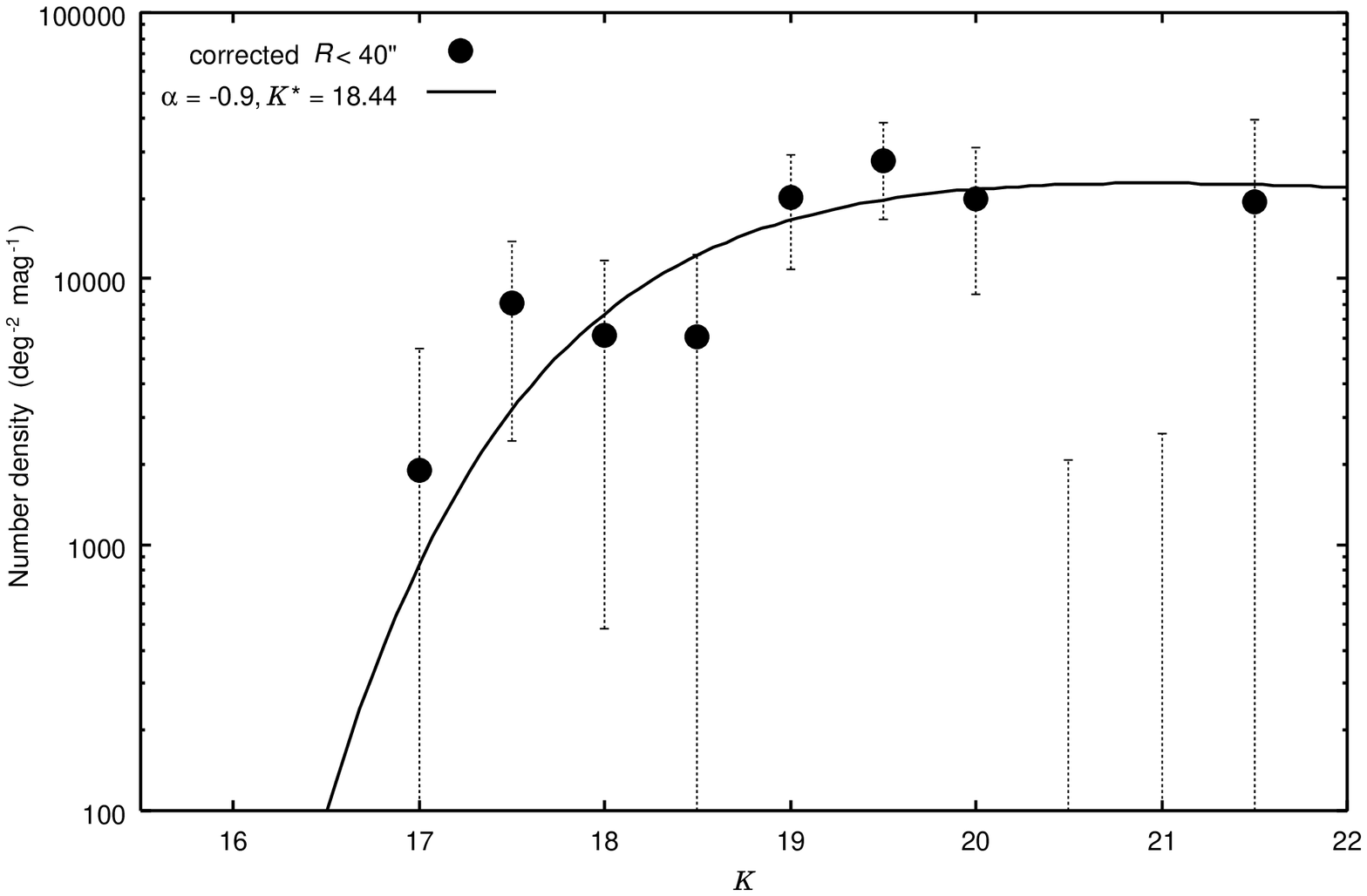,scale=0.65}
 \end{center}
\footnotesize Fig.\ 7.\
Luminosity function of the cluster region. The solid line is the Schechter function fitted to the points between $K=17$ and 20 mag. Note that the observed values at $K=20.5$ and 21.0 are negative.
\end{figure*}

\begin{figure*}[p]
\begin{center}
   \epsfile{file=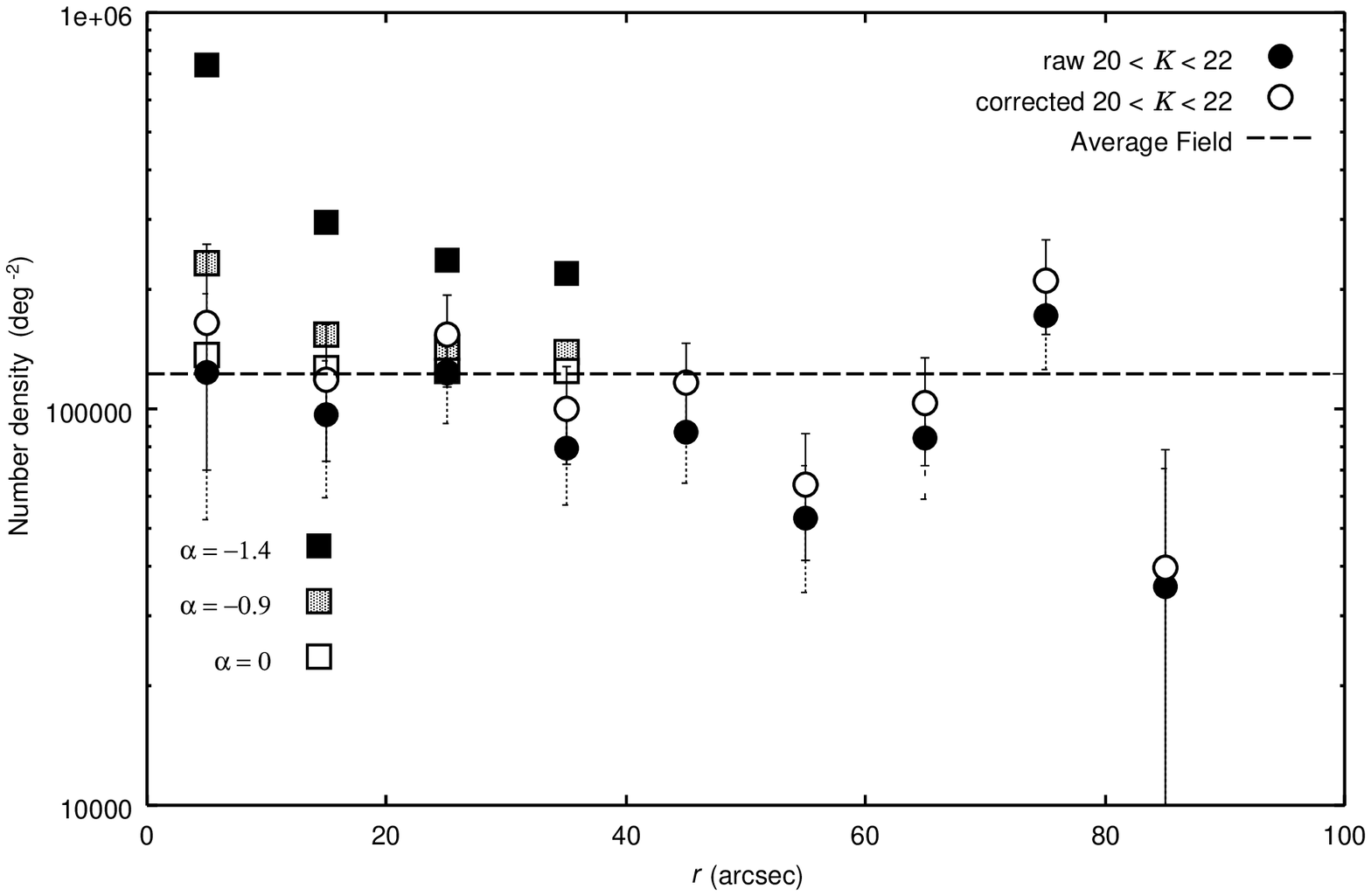,scale=0.6}
 \end{center}
\footnotesize Fig.\ 8.\
Raw (filled circles) and the corrected (open circles) surface
number density profile of the galaxies with 20 $< K <$ 22 in function
of the radius from 3C 324.  The dashed line represents the averaged
surface number densities in the field shown in figure 2. For a 
comparison, we also show the expected profiles derived from the counts
at $K=$ 17--20 mag assuming the Schecter function and various values of
the faint-end slope (squares). 
\end{figure*}

\begin{figure*}[p]
\begin{center}
   \epsfile{file=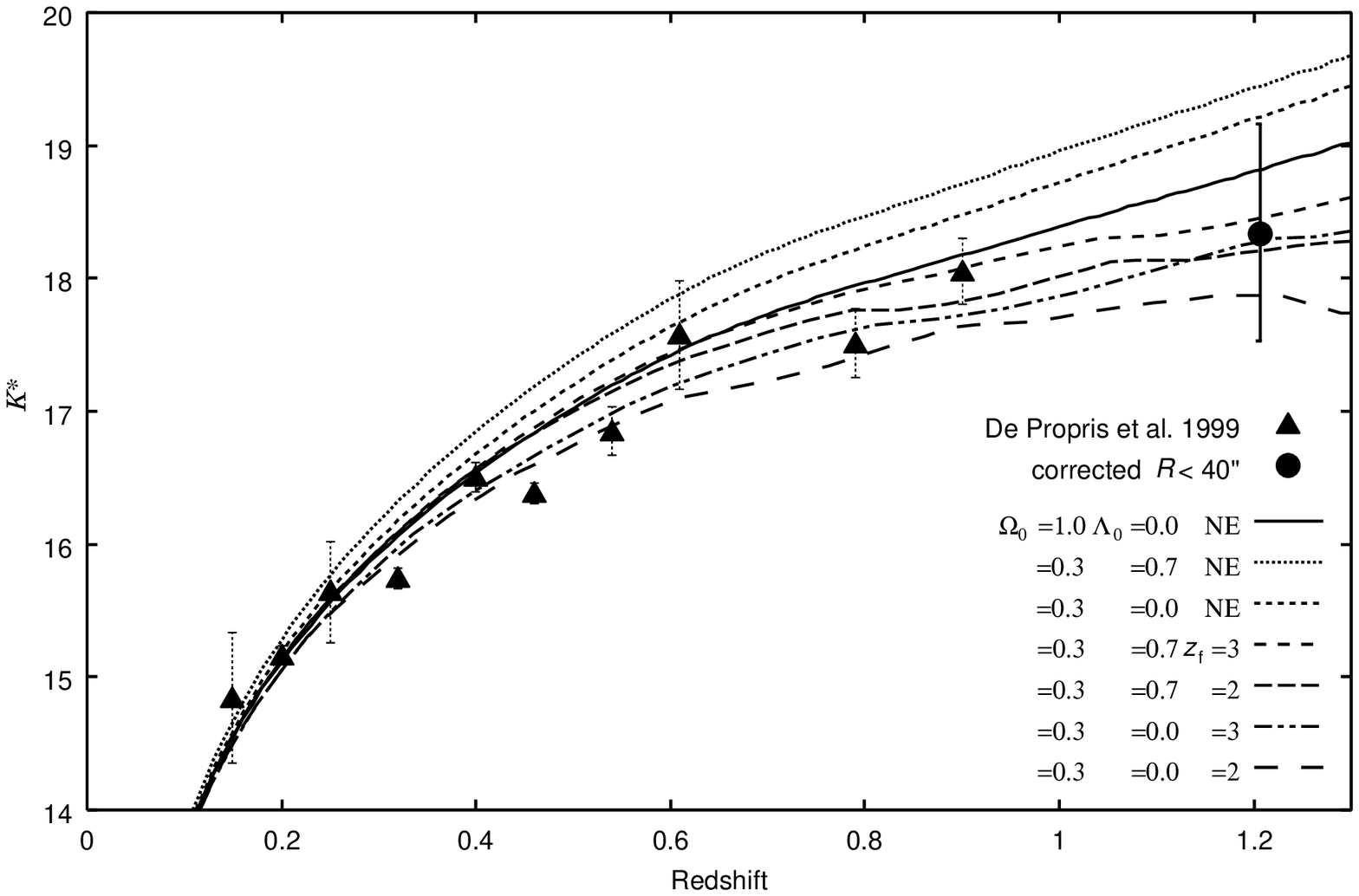,scale=0.7}
 \end{center}
\footnotesize Fig.\ 9.\
$K^{*}$-$z$ Hubble diagram for the  clusters at $z=0.1-0.9$ studied
by De Propris et al. (1999) and the 3C 324 cluster.  Lines represent
galaxy models calculated using GISELL96 (see text) under the various
set of the cosmological parameters, $\Omega_0$ and $\Lambda_0$.
H$_{0} =$ 65 km s$^{-1}$ Mpc$^{-1}$ is adopted.  
\end{figure*}

\end{document}